\documentclass[12pt,letterpaper]{article}

\title{Charged rotating black holes in six-dimensional gauged supergravity}
\author{
David D. K. Chow
}
\date{}

\usepackage{amsmath}
\usepackage{amsfonts}
\usepackage{amssymb}
\usepackage{amscd}
\usepackage{latexsym}

\makeatletter
\@addtoreset{equation}{section}
\makeatother

\newcommand{\ben}{\begin{equation}}
\newcommand{\een}{\end{equation}}
\newcommand{\bea}{\setlength\arraycolsep{2pt} \begin{eqnarray}}
\newcommand{\eea}{\end{eqnarray}}
\newcommand{\nnr}{\nonumber \\}
\newcommand{\pd}{\partial}
\newcommand{\ud}{\textrm{d}}

\newcommand{\ue}{\textrm{e}}
\newcommand{\uU}{\textrm{U}}
\newcommand{\uSO}{\textrm{SO}}
\newcommand{\uSU}{\textrm{SU}}
\newcommand{\uAdS}{\textrm{AdS}}

\begin{document}
\begin{titlepage}
\begin{flushright}
DAMTP-2008-73
\end{flushright}
\vspace*{100pt}
\begin{center}
{\bf \Large{Charged rotating black holes in six-dimensional gauged supergravity}}\\
\vspace{50pt}
\long\def\symbolfootnote[#1]#2{\begingroup
\def\thefootnote{\fnsymbol{footnote}}\footnote[#1]{#2}\endgroup}
\large{David D. K. Chow\symbolfootnote[1]{Present address: George P. and Cynthia W. Mitchell Institute for Fundamental Physics and Astronomy, Texas A\&M University, College Station, TX 77843-4242, USA}}
\end{center}
\begin{center}
{\it Department of Applied Mathematics and Theoretical Physics, University of Cambridge,\\
Centre for Mathematical Sciences, Wilberforce Road, Cambridge CB3 0WA, UK}\\
{\tt D.D.K.Chow@damtp.cam.ac.uk}\\
\vspace{50pt}
{\bf Abstract\\}
\end{center}
We obtain non-extremal charged rotating black holes in six-dimensional $\uSU (2)$ gauged
supergravity with two independent angular momenta and one $\uU (1)$ charge.  These include supersymmetric black holes without naked closed timelike curves.
\end{titlepage}

\newpage

\section{Introduction}

Black hole solutions of gauged supergravity theories are of interest because they provide gravitational backgrounds for studying the
AdS/CFT correspondence \cite{AdSCFT, gaugecorncst, AdShol} (see \cite{AdSCFTPhysRep} for a review), which relates string theory or M-theory on $\uAdS_D \times X$, for some compact manifold $X$, to a $(D-1)$-dimensional conformal field theory on the boundary of AdS.  Numerous charged
and rotating AdS black holes in various dimensions have been constructed over the past few years; for a review, see for example \cite{bhhd, equalcharge}.  These have focussed on supergravities in spacetime dimensions $D = 4, 5, 7$.  There are examples of both supersymmetric and non-supersymmetric black holes in all of these dimensions.  All known solutions make the simplification to abelian gauge fields in the Cartan
subgroup, which is of the form $\uU (1)^k$, of the full gauge group.  Furthermore, these solutions all employ one of three additional simplifying strategies: setting some of the $\uU (1)$ charges equal, setting all of
the angular momenta equal, or restricting to supersymmetric
solutions.  In these dimensions, a black hole with all angular momenta and $\uU (1)$ charges independent is not currently known.

Six dimensions is a slightly different case.  There is an $\mathcal{N} = 4$,
$\uSU (2)$ gauged supergravity theory that can be obtained from
dimensional reduction of massive type IIA supergravity on $S^4$ (more
precisely, a hemisphere of $S^4$) \cite{6dsugraIIA}.  The theory has
half-maximal supersymmetry and admits an $\uAdS _6$ vacuum that has a
ten-dimensional interpretation as the near-horizon limit of a
localized D4--D8 brane configuration.  Through the AdS/CFT
correspondence, the theory is expected to be dual to a $D = 5$,
$\mathcal{N} = 2$ superconformal field theory.

Here, we obtain the first charged and rotating AdS black holes of this gauged supergravity theory.  The solution has two independent angular momenta and one independent $\uU (1)$ charge, i.e.~the maximum number possible.  We compute the thermodynamic quantities and find that there are supersymmetric solutions.  There are supersymmetric black holes without naked closed timelike curves; these are the first known supersymmetric AdS$_6$ black holes.

\section{Black hole solution}

The bosonic fields of $D = 6$, $\mathcal{N} = 4$, $\uSU (2)$ gauged
supergravity are a graviton, a two-form potential, a one-form
potential, the gauge potentials of $\uSU (2)$ Yang--Mills theory, and
one scalar.   The bosonic Lagrangian is
\bea
\mathcal{L}_6 & = & R \star 1 - \frac{1}{2} \star \ud \varphi \wedge
\ud \varphi - \frac{1}{2} X^{-2} ( \star F_{(2)} \wedge F_{(2)} + \star F_{(2)}^I \wedge F_{(2)}^I ) -
\frac{1}{2} X^4 \star F_{(3)} \wedge F_{(3)} \nnr
&& + g^2 (9 X^2 + 12 X^{-2} - X^{-6}) \star 1 \nnr
&& - A_{(2)} \wedge \left( \frac{1}{2} \ud A_{(1)} \wedge \ud A_{(1)}
  + \frac{g}{\sqrt{2}} A_{(2)} \wedge \ud A_{(1)} + \frac{g^2}{3}
  A_{(2)} \wedge A_{(2)} + \frac{1}{2} F_{(2)}^I \wedge
  F_{(2)}^I \right) ,
\label{Lagrangian}
\eea
where $F_{(2)} = \ud A_{(1)} + \sqrt{2} g A_{(2)}$, $F_{(2)}^I = \ud
A_{(1)}^I + (3 / \sqrt{2}) g \epsilon_{IJK} A_{(1)}^J \wedge
A_{(1)}^K$, $F_{(3)} = \ud A_{(2)}$, and $X = \ue^{- \varphi /
  \sqrt{8}}$.

The resulting Einstein equation is
\bea
G_{ab} & = & \frac{1}{2} \nabla_a \varphi \, \nabla_b \varphi -
\frac{1}{4} \nabla^c \varphi \, \nabla_c \varphi \, g_{ab} + X^{-2}
\left( \frac{1}{2} F{_a}{^c} F_{bc} - \frac{1}{8} F^{cd} F_{cd} g_{ab}
\right) \nnr
&& + X^{-2} \left( \frac{1}{2} F{^I}{_a}{^c} F{^I}{_{bc}}
  - \frac{1}{8} F^{Icd}F{^I}{_{cd}} g_{ab} \right) + X^4 \left(
  \frac{1}{4} F{_a}^{cd} F_{bcd} - \frac{1}{24} F^{cde} F_{cde} g_{ab}
\right) \nnr
&& + \frac{g^2}{2} (9 X^2 + 12 X^{-2} - X^{-6}) g_{ab} .
\eea
The remaining field equations are
\bea
&& \square \varphi = \frac{1}{\sqrt{8}} X^{-2} ( F^{ab} F_{ab} + F^{Iab} F{^I}{_{ab}} )  - \frac{1}{3 \sqrt{8}} X^4 F^{abc}
F_{abc} + \frac{3}{\sqrt{2}} g^2 (3 X^2 - 4 X^{-2} + X^{-6}) , \nnr
&& \ud (X^{-2} \star F_{(2)}) = - F_{(2)} \wedge F_{(3)} , \nnr
&& \ud (X^{-2} \star F_{(2)}^I) + \frac{3 g}{\sqrt{2}} \epsilon_{IJK}
X^{-2} A_{(1)}^J \wedge \star F_{(2)}^K = - F_{(2)}^I \wedge F_{(3)} ,
\nnr
&& \ud (X^4 \star F_{(3)}) = - \frac{1}{2} F_{(2)} \wedge F_{(2)} -
\frac{1}{2} F_{(2)}^I \wedge F_{(2)}^I - \sqrt{2} g X^{-2} \star F_{(2)} .
\eea

We shall truncate to the sector with $A_{(1)}^1 = A_{(1)}^2 = 0$, so that
only a single $\uU (1)$ field of the $\uSU (2)$ gauge group is
excited: $A_{(1)}^3 \neq 0$.  By a gauge transformation $A_{(2)} \rightarrow A_{(2)} + \ud \Lambda_{(1)}$, $A_{(1)} \rightarrow A_{(1)} - \sqrt{2} g \Lambda_{(1)}$, which is possible for $g \neq 0$, we shall also take $A_{(1)} = 0$.  We
then rescale and relabel $A_{(1)}^3 \rightarrow \sqrt{2} A_{(1)}$, so then $F_{(2)}^3 \rightarrow
\sqrt{2} F_{(2)}$.  The bosonic
field equations can be obtained from the Lagrangian
\bea
\mathcal{L}_6 & = & R \star 1 - \frac{1}{2} \star \ud \varphi \wedge
\ud \varphi - X^{-2} (\star F_{(2)} \wedge F_{(2)} + g^2 \star A_{(2)}
\wedge A_{(2)}) - \frac{1}{2} X^4 \star F_{(3)} \wedge F_{(3)} \nnr
&& + g^2 (9 X^2 + 12
X^{-2} - X^{-6}) \star 1 - F_{(2)} \wedge F_{(2)} \wedge A_{(2)} -
\frac{g^2}{3} A_{(2)}
\wedge A_{(2)} \wedge A_{(2)} ,
\eea
where $F_{(2)} = \ud A_{(1)}$.

A charged and rotating black hole solution is
\bea
&& \ud s^2 = H^{1/2} \bigg[ \frac{(r^2 + y^2) (r^2 + z^2)}{R} \ud r^2
+ \frac{(r^2 + y^2) (y^2 - z^2)}{Y} \ud y^2 + \frac{(r^2 + z^2)
  (z^2 - y^2)}{Z} \ud z^2 \nnr
&& \qquad \qquad \quad - \frac{R}{H^2 (r^2 + y^2) (r^2 + z^2)} \mathcal{A}^2 \nnr
&& \qquad \qquad \quad + \frac{Y}{(r^2 + y^2) (y^2 - z^2)} \left( \ud t' +
  (z^2 - r^2) \, \ud \psi_1 - r^2 z^2 \, \ud \psi_2 - \frac{qr \mathcal{A}}{H (r^2 +
    y^2) (r^2 + z^2)} \right) ^2 \nnr
&& \qquad \qquad \quad + \frac{Z}{(r^2 + z^2) (z^2 - y^2)} \left( \ud t' +
  (y^2 - r^2) \, \ud \psi_1 - r^2 y^2 \, \ud \psi_2 - \frac{qr \mathcal{A}}{H (r^2 +
    y^2) (r^2 + z^2)} \right) ^2 \bigg] , \nnr
&& X = H^{-1/4} , \quad A_{(1)} = \frac{2 m s c r}{H (r^2 + y^2) (r^2 +
  z^2)} \mathcal{A} , \nnr
&& A_{(2)} = \frac{q}{H (r^2 + y^2)^2 (r^2 + z^2)^2} \bigg[ -
\frac{yz [ 2 r (2 r^2 + y^2 + z^2) + q ]}{H} \ud r \wedge \mathcal{A}
\nnr
&& \qquad \qquad \qquad \qquad \qquad \qquad + z [ (r^2 + z^2) (r^2 - y^2) + q r ] \, \ud y \nnr
&& \qquad \qquad \qquad \qquad \qquad \qquad \quad \wedge \left( \ud t' + (z^2 - r^2) \, \ud \psi_1
  - r^2 z^2 \, \ud \psi_2 - \frac{qr \mathcal{A}}{H (r^2 + y^2) (r^2 + z^2)}
  \right) \nnr
&& \qquad \qquad \qquad \qquad \qquad \qquad + y [ (r^2 + y^2) (r^2 - z^2) + q r ] \, \ud z \nnr
&& \qquad \qquad \qquad \qquad \qquad \qquad \quad \wedge \left( \ud t' + (y^2 - r^2) \, \ud \psi_1
  - r^2 y^2 \, \ud \psi_2 - \frac{qr \mathcal{A}}{H (r^2 + y^2) (r^2 + z^2)}
  \right) \bigg] , \nnr
\eea
where
\bea
&& R = (r^2 + a^2) (r^2 + b^2) + g^2 [ r (r^2 + a^2) + q ] [ r (r^2 +
b^2) + q ] - 2 m r , \nnr
&& Y = - (1 - g^2 y^2) (a^2 - y^2) (b^2 - y^2) , \quad Z = - (1 - g^2
z^2) (a^2 - z^2) (b^2 - z^2), \nnr
&& H = 1 + \frac{qr}{(r^2 + y^2) (r^2 + z^2)} , \quad q = 2 m s^2 ,
\quad s = \sinh \delta , \quad c = \cosh \delta , \nnr
&& \mathcal{A} = \ud t' + (y^2 + z^2) \, \ud \psi_1 + y^2 z^2 \, \ud \psi_2 .
\eea

We have presented the solution most concisely by using Jacobi--Carter
coordinates.  To compute thermodynamic quantities, we use angular velocities measured with respect to a non-rotating frame at infinity and move to Boyer--Lindquist time and azimuthal coordinates $t, \phi_1, \phi_2$ through the coordinate change
\ben
t' = \tilde{t} - a^4 \tilde{\phi}_1 - b^4 \tilde{\phi}_2 , \quad
\psi_1 = - g^2 \tilde{t} + a^2 \tilde{\phi}_1 + b^2 \tilde{\phi}_2
, \quad \psi_2 = g^4 \tilde{t} - \tilde{\phi}_1 - \tilde{\phi}_2
,
\een
where
\ben
\tilde{t} = \frac{t}{\Xi_a \Xi_b} , \quad \tilde{\phi}_1 =
\frac{\phi_1}{\Xi_a a (a^2 - b^2)} , \quad \tilde{\phi}_2 =
\frac{\phi_2}{\Xi_b b (b^2 - a^2)} ,
\een
with $\Xi_a = 1 - a^2 g^2$ and $\Xi_b = 1 - b^2 g^2$.  $\phi_1$ and
$\phi_2$ are canonically normalized, with period $2 \pi$, and $t$ is also canonically normalized.  The other coordinates are $r$ as the usual radial coordinate, and $y$ and $z$ representing the latitudinal coordinates.  The ranges of $y$ and $z$ can be taken as $-a \leq y \leq a \leq z \leq b$, or with appropriate permutations and sign changes.  $y$ and $z$ are useful coordinates for describing the round metric on $S^4$, because they parameterize direction cosines in a rather symmetric manner \cite{genKNUTAdSalld}.  One could replace $y$ and $z$ with latitudinal spherical polar coordinates to see that singularities at $y = z$ and when $a = b$ can be removed, and that the gauge potential is globally defined.

The solution has five parameters: a mass parameter $m$; two angular
rotation parameters, $a$ and $b$, that describe rotation in orthogonal
2-planes; a charge parameter $\delta$; and a gauge-coupling constant
$g$.  It is the most general black hole solution of this six-dimensional $\uSU (2)$ gauged supergravity theory that is known.  More general gauge fields would require non-abelian gauge fields.

In the absence of rotation, with $a = b = 0$, our
solution reduces to the static solution of \cite{6dsugraIIA}.
In the absence of charge, with $\delta = 0$, our solution reduces to
the six-dimensional Kerr--AdS metric \cite{genKerrdS, rotbhhigherdim},
as presented in \cite{genKNUTAdSalld}.  In the absence of
gauging, with $g = 0$, our solution reduces to the two-charge
Cveti\v{c}--Youm solution \cite{nearBPSsat}, but with both charges equal
and presented, up to a gauge transformation of $A_{(2)}$, as in \cite{equalcharge}.

To find this solution, we have been particularly helped by the structure of similar gauged supergravity solutions in other dimensions, as summarized in \cite{equalcharge}.  As mentioned, one difference to note is that the two-form potential $A_{(2)}$ here does not coincide with the $(D-4)$-form potential $A_{(D-4)}$ presented in \cite{equalcharge}.  However, the two potentials differ by an exact form, so give rise to the same field strength.  If desired, one can perform a gauge transformation so that the two potentials coincide and then, in the notation of (\ref{Lagrangian}), we have non-vanishing $A_{(1)}$.

\section{Thermodynamics}

The outer black hole horizon is located at the largest root of $R(r)$,
say at $r = r_+$.  Its angular velocities, $\Omega_a$ and $\Omega_b$, are constant over the
horizon and are obtained from the Killing vector
\ben
\ell = \frac{\pd}{\pd t} +
\Omega_a \frac{\pd}{\pd \phi_1} + \Omega_b \frac{\pd}{\pd \phi_2}
\een
that becomes
null on the horizon.  The angular momenta, $J_a$ and $J_b$, are given by the Komar
integrals
\ben
J_i = \frac{1}{16 \pi} \int_{S^4_\infty} \, \star \, \ud K_i ,
\een
where $K_i$ is the one-form obtained from the Killing vector $\pd / \pd
\phi_i$.  The electrostatic potential, which is also constant over the
horizon, is $\Phi = \ell \cdot A_{(1)} | _{r = r_+}$.  The conserved
electric charge is, for our normalization,
\ben
Q = \frac{1}{8 \pi} \int_{S^4_\infty} \, (X^{-2} \star F_{(2)} + F_{(2)}
\wedge A_{(2)}) ,
\een
although for our solution only $\star F_{(2)}$ contributes.  The
horizon area $A$ is obtained by integrating the square root of the
determinant of the induced metric on a time slice of the horizon,
\ben
\det g_{( y, z, \phi_1, \phi_2 )} |_{r = r_+} = \frac{[(r_+^2 + a^2)
  (r_+^2 + b^2) + q r_+]^2 (y^2 - z^2)^2}{\Xi_a^2 \Xi_b^2 a^2 b^2 (a^2
  - b^2)^2} .
\een
The surface gravity $\kappa$, again constant over the horizon, is
given by $\ell^b \, \nabla_b l^a = \kappa \ell^a$ evaluated on the
horizon.  As usual, we take the temperature to be $T = \kappa / 2 \pi$ and the
entropy to be $S = A / 4$.

One finds that $T \, \ud S + \Omega_a \, \ud J_a + \Omega_b \, \ud J_b + \Phi
\, \ud Q$ is an exact differential, and so we may integrate the first law
of black hole mechanics,
\ben
\ud E = T \, \ud S + \Omega_a \, \ud J_a + \Omega_b \, \ud J_b + \Phi
\, \ud Q ,
\een
to obtain an expression for the thermodynamic mass $E$.  There are
other methods of computing the mass of AdS black holes: see, for example, \cite{1stlaw, massrotbh}.

In summary, we find the thermodynamic quantities
\bea
E & = & \frac{\pi}{3 \Xi_a \Xi_b} \left[ 2m \left( \frac{1}{\Xi_a} +
    \frac{1}{\Xi_b} \right) + q \left( 1 + \frac{\Xi_a}{\Xi_b} +
    \frac{\Xi_b}{\Xi_a} \right) \right] , \nnr
S & = & \frac{2 \pi^2
  [(r_+^2 + a^2) (r_+^2 + b^2) + q r_+]}{3 \Xi_a \Xi_b} , \nnr
T & = & \frac{2 (1 + g^2 r_+^2) r_+^2 (2 r_+^2 + a^2 + b^2) - (1 - g^2
  r_+^2) (r_+^2 + a^2) (r_+^2 + b^2) + 4 q g^2 r_+^3 - q^2 g^2}{4 \pi
  r_+ [(r_+^2 + a^2) (r_+^2 + b^2) + q r_+]} , \nnr
J_a & = & \frac{2 \pi m a (1 + \Xi_b s^2)}{3 \Xi_a^2 \Xi_b} ,
\quad \Omega_a = \frac{a [(1 + g^2 r_+^2) (r_+^2 + b^2) + q g^2
  r_+]}{(r_+^2 + a^2) (r_+^2 + b^2) + q r_+} , \nnr
J_b & = & \frac{2 \pi m b (1 + \Xi_a s^2)}{3 \Xi_a \Xi_b^2} , \quad
\Omega_b = \frac{b [(1 + g^2 r_+^2) (r_+^2 + a^2) + q g^2 r_+]}{(r_+^2
  + a^2) (r_+^2 + b^2) + q r_+} , \nnr
Q & = & \frac{2 \pi m s c}{\Xi_a \Xi_b} , \quad \Phi = \frac{2 m s c
  r_+}{(r_+^2 + a^2) (r_+^2 + b^2) + q r_+} .
\eea

\section{Supersymmetric solutions}

The algebra of the supercharges $\mathcal{Q}$ in $D = 6$, $\mathcal{N}
= 4$ gauged supergravity is $\{ \mathcal{Q} , \mathcal{Q} \} = \{
\overline{\mathcal{Q}} , \overline{\mathcal{Q}} \} = 0$, and
\ben
M := \{ \mathcal{Q} , \overline{\mathcal{Q}} \} = \frac{1}{2} J_{AB} \gamma^{AB} + Z ,
\een
where $\gamma^{AB}$ are generators of $\uSO (5,2)$, the supercharges
$\mathcal{Q}$ are 8-component Dirac spinors of $\uSO (5,2)$, and $Z$
is the central charge.  We take
\ben
J_{06} = g^3 E , \quad J_{12} = g^4 J_a , \quad J_{34} = g^4 J_b , \quad Z = g^3 Q ,
\een
so the eigenvalues of the Bogomolny matrix $g^{-3} M$ are
\ben
\lambda = E \pm J_a \pm J_b \pm Q .
\een
After a choice of signs, we may take the BPS condition to be
\ben
E - g J_a - g J_b - Q = 0 ,
\een
which is satisfied if
\ben
\ue^{2 \delta} = 1 + \frac{2}{(a+b) g} ,
\label{BPSsol}
\een
or equivalently
\ben
q = \frac{2 m}{(a + b) g (2 + ag + bg)} .
\een
Such supersymmetric solutions generally preserve $1/8$ supersymmetry.

Imposing the BPS condition, the radial function in the metric is
\ben
R = [g r^3 - (a + b + abg) r + qg]^2 + [(1 + ag + bg) r^2 - ab]^2 ,
\een
which is a sum of two squares.  At a horizon, $R = 0$, and so there
are supersymmetric black holes if
\ben
q = \frac{\Xi_{a+} \Xi_{b+} (a + b) r_+}{(1 + ag + bg) g} ,
\label{susybhq}
\een
where
\ben
r_+ = \sqrt{\frac{ab}{1 + ag + bg}} ,
\label{rplus}
\een
and we have denoted $\Xi_{a+} = 1 + ag$ and $\Xi_{b+} = 1 + bg$.
Being supersymmetric, these black holes necessarily have zero
temperature.

An alternative way of singling out these supersymmetric black holes from the supersymmetric solutions is
to consider whether or not the spacetime suffers from the pathology of
naked closed timelike curves (CTCs).  We can write the metric in the form
\bea
\ud s^2 & = & H^{1/2} \bigg[ \frac{(r^2 + y^2) (r^2 + z^2)}{R} \ud r^2
+ \frac{(r^2 + y^2) (y^2 - z^2)}{Y} \ud y^2 + \frac{(r^2 + z^2) (z^2 -
  y^2)}{Z} \ud z^2 \nnr
&& \qquad \quad - \frac{RYZ}{H^2 \Xi_a^2 \Xi_b^2 a^2 b^2 (a^2 - b^2)^2 B_1 B_2} \ud t^2 + B_2 (\ud
\phi_2 + v_{21} \, \ud \phi_1 + v_{20} \, \ud t)^2 \nnr
&& \qquad \quad + B_1 (\ud \phi_1 +
v_{10} \, \ud t)^2 \bigg] ,
\eea
so that the periodic $\phi_i$ coordinates have been separated from a
$\ud t^2$ term.  There are closed timelike curves if any of $B_1$ or $B_2$ are negative.  We do not, at this stage, require any detailed
knowledge of the additional functions introduced, which can be
straightforwardly computed; we have only used the fact that there is a simple expression for $\det (g_{ab})$.  For supersymmetric solutions, the Killing vector
\ben
K = \frac{\pd}{\pd t} + g \frac{\pd}{\pd \phi_1} + g \frac{\pd}{\pd \phi_2}
\een
is the square of a Killing spinor $\epsilon$, i.e.~$K^a =
\bar{\epsilon} \gamma^a \epsilon$.  If we define $K =: \pd / \pd \bar{t}$, and transform
from the $t$ coordinate in favour of $\bar{t}$, then $\ud \phi_i \rightarrow \ud \phi_i + g \, \ud \bar{t}$.  The $g_{\bar{t} \bar{t}}$ component
of the metric then gives
\bea
K^a K_a & = & H^{1/2} \bigg[ - \frac{RYZ}{H^2 \Xi_a^2 \Xi_b^2 a^2 b^2
  (a^2 - b^2)^2 B_1 B_2} + B_2 [ g (1 + v_{21}) + v_{20} ]^2 + B_1 (g +
v_{10} )^2 \bigg] . \nnr
\label{gtt}
\eea
Because of its spinorial square
root, one can show from Fierz identities that $K$ is non-spacelike,
and its norm can be expressed involving a sum of squares.  In this
case, we have
\bea
K^a K_a & = & - H^{-3/2} \bigg[ \left( 1 - \frac{(1 + ag + bg) q g
    r}{\Xi_{a+} \Xi_{b+} (a + b) r_{\textrm{h}}^2}
  \frac{(r_{\textrm{h}}^2 + y^2) (r_{\textrm{h}}^2 + z^2)}{(r^2 + y^2)
    (r^2 + z^2)} \right) ^2 \nnr
&& \qquad \quad + \frac{g^2 y^2 z^2}{a^2 b^2} \left( r - \frac{(1 + ag + bg)
    qg}{\Xi_{a+} \Xi_{b+} (a + b)} \frac{(r_{\textrm{h}}^2 + y^2)
    (r_{\textrm{h}}^2 + z^2)}{(r^2 + y^2) (r^2 + z^2)} + \frac{q (r^2
    - r_{\textrm{h}}^2)}{(r^2 + y^2) (r^2 + z^2)} \right) ^2 \bigg] , \nnr
\eea
where
\ben
r_{\textrm{h}}^2 = \frac{ab}{1 + ag + bg} .
\een
At a horizon, $R = 0$, and so, since the left hand side of (\ref{gtt}) is not positive there, in general $B_1$ or $B_2$ will be negative near a horizon, resulting in naked closed timelike curves.  The exception is if the left
hand side vanishes at some radius $r = r_+$.  Then we must have $r_+
= r_{\textrm{h}}$, in agreement with (\ref{rplus}), and then
(\ref{susybhq}) must also hold, so we have the supersymmetric black
hole.  Hence there are no CTCs just outside the horizon, and, if furthermore the parameters satisfy certain (complicated) inequalities, then there are
no CTCs anywhere outside the horizon.  For example, for arbitrary positive $a$
and $b$, taking $g$ to be positive and small ensures that there are no
naked closed timelike curves.

For these supersymmetric black holes, it is not possible for
additional eigenvalues of the Bogomolny matrix to vanish without
violating certain conditions: $-1 < ag < 1$ and $-1 < bg < 1$, for the
correct signature; and $(a + b) g > 0$, to satisfy (\ref{BPSsol}).
Therefore all supersymmetric black holes preserve $1/8$
supersymmetry.

In common with all other dimensions studied in the literature, we have found supersymmetric black holes in six-dimensional gauged supergravity; see references within \cite{bhhd} for supersymmetric black holes in other dimensions.  However, in $D = 5$ and $D = 7$, there are also supersymmetric topological soliton solutions, which have a geometry that ends at some
minimum radius $r = r_0$.  In particular, see \cite{rotbhgaugedsugra}; however, it has been argued that these are not regular \cite{nsusysmooth}.  Here, in $D = 6$, demanding a supersymmetric solution with
zero horizon area gives $r_0^2 = - a^2 b^2 g^2 / (1 + ag + bg)^2$.  In
even dimensions the radial coordinate $r$ is real, whereas in odd
dimensions one may continue $r^2$ to negative values, and so here in
$D = 6$ no supersymmetric topological soliton is possible.  Similarly, in $D = 4$, the other even dimension that has been studied, supersymmetric topological solitons were not found in \cite{solbh, rotbhgaugedsugra}.

\section{Symmetries and separability}

Because the metric falls into the family considered in
\cite{equalsym}, it follows that there are two irreducible rank-2 conformal
Killing--St\"{a}ckel tensors, that the Hamilton--Jacobi equation for
null geodesics separates, and that the massless Klein--Gordon equation
separates.  Furthermore, the conformally related metric $\ud
\tilde{s}^2 = H^{-1/2} \, \ud s^2$ possesses a separability structure,
with two irreducible rank-2 Killing--St\"{a}ckel tensors.

\section{Discussion}

We have found a non-extremal black hole solution of $D = 6$ gauged
supergravity with the maximum number of independent angular momenta
and abelian charges.  As such, it may be the unique black hole solution of the theory.  The same remains to be done for gauged supergravities in dimensions $D = 4, 5, 7$, however such solutions are expected to be rather more complicated.  Such AdS black holes would be useful for studying the AdS/CFT correspondence.

\section*{Acknowledgements}

I would like to thank Chris Pope for reading a draft of this manuscript.  This work has been supported by STFC.


\begin{thebibliography}{99}

\bibitem{AdSCFT} J. Maldacena, ``The large $N$ limit of superconformal
  field theories and supergravity,'' \emph{Adv. Theor. Math. Phys.} {\bf 2}, 231 (1998), [{\tt hep-th/9711200}].

\bibitem{gaugecorncst} S.S. Gubser, I.R. Klebanov and A.M. Polyakov,
  ``Gauge theory correlators from non-critical string theory,''
  \emph{Phys. Lett. B} {\bf 428}, 105 (1998), [{\tt hep-th/9802109}].

\bibitem{AdShol} E. Witten, ``Anti-de Sitter space and holography,''
  \emph{Adv. Theor. Math. Phys.} {\bf 2}, 253 (1998), [{\tt
    hep-th/9802150}].

\bibitem{AdSCFTPhysRep} O. Aharony, S.S. Gubser, J. Maldacena,
  H. Ooguri and Y. Oz, ``Large $N$ field theories, string theory and
  gravity,'' \emph{Phys. Rep.} {\bf 323}, 183 (2000), [{\tt
    hep-th/9905111}].

\bibitem{bhhd} R. Emparan and H.S. Reall, ``Black holes in higher
  dimensions,'' \emph{Living Rev. Rel.} {\bf 11}, 6 (2008), [{\tt arXiv:0801.3471}].

\bibitem{equalcharge} D.D.K. Chow, ``Equal charge black holes and
  seven-dimensional gauged supergravity,'' \emph{Class. Quant. Grav.} {\bf 25}, 175010 (2008), [{\tt arXiv:0711.1975}].

\bibitem{6dsugraIIA} M. Cveti\v{c}, H. L\"{u} and C.N. Pope, ``Gauged
  six-dimensional supergravity from massive type IIA string theory,''
  \emph{Phys. Rev. Lett.} {\bf 83}, 5226 (1999), [{\tt hep-th/9906221}].

\bibitem{genKNUTAdSalld} W. Chen, H. L\"{u} and C.N. Pope, ``General
  Kerr--NUT--AdS metrics in all dimensions,''
  \emph{Class. Quant. Grav.} {\bf 23}, 5323 (2006), [{\tt
    hep-th/0604125}].

\bibitem{genKerrdS} G.W. Gibbons, H. L\"{u}, D.N. Page and
  C.N. Pope, ``The general Kerr--de Sitter metrics in all
  dimensions,'' \emph{J. Geom. Phys.} {\bf 53}, 49 (2004), [{\tt hep-th/0404008}].

\bibitem{rotbhhigherdim} G.W. Gibbons, H. L\"{u}, D.N. Page and
  C.N. Pope, ``Rotating black holes in higher dimensions with a
  cosmological constant,'' \emph{Phys. Rev. Lett.} {\bf 93}, 171102
  (2004), [{\tt hep-th/0409155}].

\bibitem{nearBPSsat} M. Cveti\v{c} and D. Youm, ``Near-BPS-saturated
  rotating electrically charged black holes as string states,''
  \emph{Nucl. Phys. B} {\bf 477}, 449 (1996), [{\tt hep-th/9605051}].

\bibitem{1stlaw} G.W. Gibbons, M.J. Perry and C.N. Pope, ``The
  first law of thermodynamics for Kerr--anti-de Sitter black holes,''
  \emph{Class. Quant. Grav.} {\bf 22}, 1503 (2005), [{\tt
    hep-th/0408217}].

\bibitem{massrotbh} W. Chen, H. L\"{u} and C.N. Pope, ``Mass of
  rotating black holes in gauged supergravities,''
  \emph{Phys. Rev. D} {\bf 73}, 104036 (2006), [{\tt
    hep-th/0510081}].

\bibitem{rotbhgaugedsugra} M. Cveti\v{c}, G.W. Gibbons, H. L\"{u} and
  C.N. Pope, ``Rotating black holes in gauged supergravities;
  thermodynamics, supersymmetric limits, topological solitons and time
  machines,'' [{\tt hep-th/0504080}].

\bibitem{nsusysmooth} S.F. Ross, ``Non-supersymmetric asymptotically AdS$_5 \times S^5$ smooth geometries,'' \emph{JHEP} {\bf 0601}, 130 (2006), [{\tt hep-th/0511090}].

\bibitem{solbh} V.A. Kosteleck\'{y} and M.J. Perry, ``Solitonic black
  holes in gauged $N=2$ supergravity,'' \emph{Phys. Lett. B} {\bf
    371}, 191 (1996), [{\tt hep-th/9512222}].

\bibitem{equalsym} D.D.K. Chow, ``Symmetries of supergravity black holes,'' [{\tt arXiv:0811.1264}].

\end{thebibliography}
\end{document}